\documentclass[twocolumn,showpacs,preprintnumbers,amsmath,amssymb]{revtex4-1}

\usepackage{graphicx}
\usepackage{color,xcolor}
\usepackage{lipsum}
\usepackage{tabularx}
\usepackage{array}
\usepackage{mathrsfs}
\usepackage{amssymb}
\usepackage{amsmath}
\usepackage{cases}  
\usepackage{graphicx}
\usepackage{subfigure} 
\usepackage{dcolumn}
\usepackage{bm}
\usepackage{verbatim} 
\usepackage[dvipdfm,pdfstartview=FitH,colorlinks,linkcolor=blue,anchorcolor=blue,citecolor=blue]{hyperref}
\usepackage{amsthm}  

\usepackage{ulem}

\begin{document}

\title{Quantum hacking perceiving for quantum key distribution using temporal ghost imaging}

\author{Fang-Xiang Wang${}^{1,2,*}$}
\author{Juan Wu${}^{1,2,*}$}
\author{Wei Chen${}^{1,2,\dagger}$}
\author{Shuang Wang${}^{1,2}$}
\author{De-Yong He${}^{1,2}$}
\author{Zhen-Qiang Yin${}^{1,2}$}
\author{Chang-Ling Zou${}^{1,2}$}
\author{Guang-Can Guo${}^{1,2}$}
\author{Zheng-Fu Han${}^{1,2}$}
\affiliation{${}^1$CAS Key Laboratory of Quantum Information, University of Science and Technology of China, Hefei 230026, China\\
${}^2$CAS Center For Excellence in Quantum Information and Quantum Physics, University of Science and Technology of China, Hefei, Anhui 230026, China\\
${}^*$These authors contributed equally to this work\\
${}^\dagger$Corresponding author: weich@ustc.edu.cn
}


\begin{abstract}
	
Quantum key distribution (QKD) can generate secure key bits between remote users with quantum mechanics. However, the gap between the theoretical model and practical realizations gives eavesdroppers opportunities to intercept secret key. The most insidious attacks, known as quantum hacking, are the ones with no significant discrepancy of the measurement results using side-channel loopholes of QKD systems. Depicting full-time-scale characteristics of the quantum signals, the quantum channel, and the QKD system can provide legitimate users extra capabilities to defeat malicious attacks. For the first time, we propose the method exploring temporal ghost imaging (TGI) scheme to perceive quantum hacking with temporal fingerprints and experimentally verify its validity. The scheme presents a common approach to promote QKD's practical security from a new perspective of signals and systems.

\end{abstract}



\maketitle


\section{Introduction}

Quantum key distribution (QKD) offers an information-theoretically secure solution for remote users to share key streams exploring the laws of quantum physics\cite{Bennett2014, Gisin2002}. However, the security of the protocol is proven based on ideal assumptions, which are hard to fit exactly in practical QKD systems. The non-ideal features in QKD systems may annihilate details of the quantum states and quantum devices' response function and compromise their practical security \cite{Scarani2009, Xu2019}. Especially, some deficiencies leave side-channel loopholes for an adversary to intercept the quantum bits without changing the dominant measurement results, and thus can not be perceived in normal QKD processing. These most threatening attacks are named as quantum hacking \cite{Makarov2005a,Qi2005,Zhao2008,Lydersen2010,Huang2013,Wiechers2011,Gerhardt2011,Bugge2014,Qian2018}.


How to bridge the security gap between the ideal model and the system is vital when developing real-life QKD systems. The ultimate solution is to design QKD protocols with minimum assumptions. For example, measurement-device-independent (MDI) \cite{Lo2012a} and twin-field QKD \cite{Lucamarini2018} protocols have successfully closed the detector-side loopholes in principle. However, these protocols are still facing technical challenges for practical applications \cite{Xu2019}, and their key rates are inferior to the widely used BB84 QKD systems up to date \cite{Bennett2014, Yuan2018}. Therefore, to improve the real-life security of prepare-and-measurement QKD systems is a valuable task. Besides optimizing the method to calculate the final secure key rate, practical QKD systems usually adopt pragmatical strategies to compete with the interceptors. For example, in order to discover the device-control attack for single-photon detectors (SPDs) \cite{Hadfield2009,Eisaman2011}, a QKD system can monitor the parameters of the SPDs, such as the photocurrent \cite{Yuan2010,Yuan2011,Elezov2019}, the optical illumination \cite{Wangs2014}, the after-pulse ratio\cite{Ferreira2012}, the backflash radiation \cite{Meda2017}, and the delayed detection events \cite{Koehler-Sidki2019}. The system can also randomly change the detection efficiency of the SPDs \cite{Wiechers2011, Lim2015, Ferreira2015, Huang2016} to perform proactive defense. 

The countermeasures mentioned above utilize the measurement results of the quantum systems, which are comprehensive and statistical. These final results may mask some essential details of the quantum signals and the system, such as the quantum devices' temporal response function. Using this kind of nonideal feature, the eavesdroppers have opportunities to hide their attack behaviors by constructing fake events with the same statistical properties but different quantum procedures. Therefore, to reveal the details of the quantum procedure will provide the extra ability for legitimate users to capture the eavesdropping.    

For the first time, we propose and demonstrate that the imaging method can contribute to perceiving quantum hacking by monitoring the quantum signals and devices. We explore the temporal ghost imaging (TGI) technique \cite{Setala2010, Shirai2010} to obtain the full-time-scale information of the quantum signals and the responding functions of the quantum systems.To verify the effectiveness of the method, we select the time-shift attack \cite{Zhao2008,Qi2005} and blinding attack \cite{Lydersen2010}, which modify the timing dimension of the quantum signals and the behaviors of the quantum devices, respectively. The proof-of-concept experiments demonstrate that TGI monitoring is a general method to defend time-domain distorting quantum hacking effectively. Furthermore, this method brings a new way to evaluate the detailed temporal information of the quantum signals and systems and is conducive for designing complex quantum information processing systems.

\section{Principle of TGI}
\label{Principle of TGI}

\begin{figure*}
	\centering
	\includegraphics[width=\textwidth]{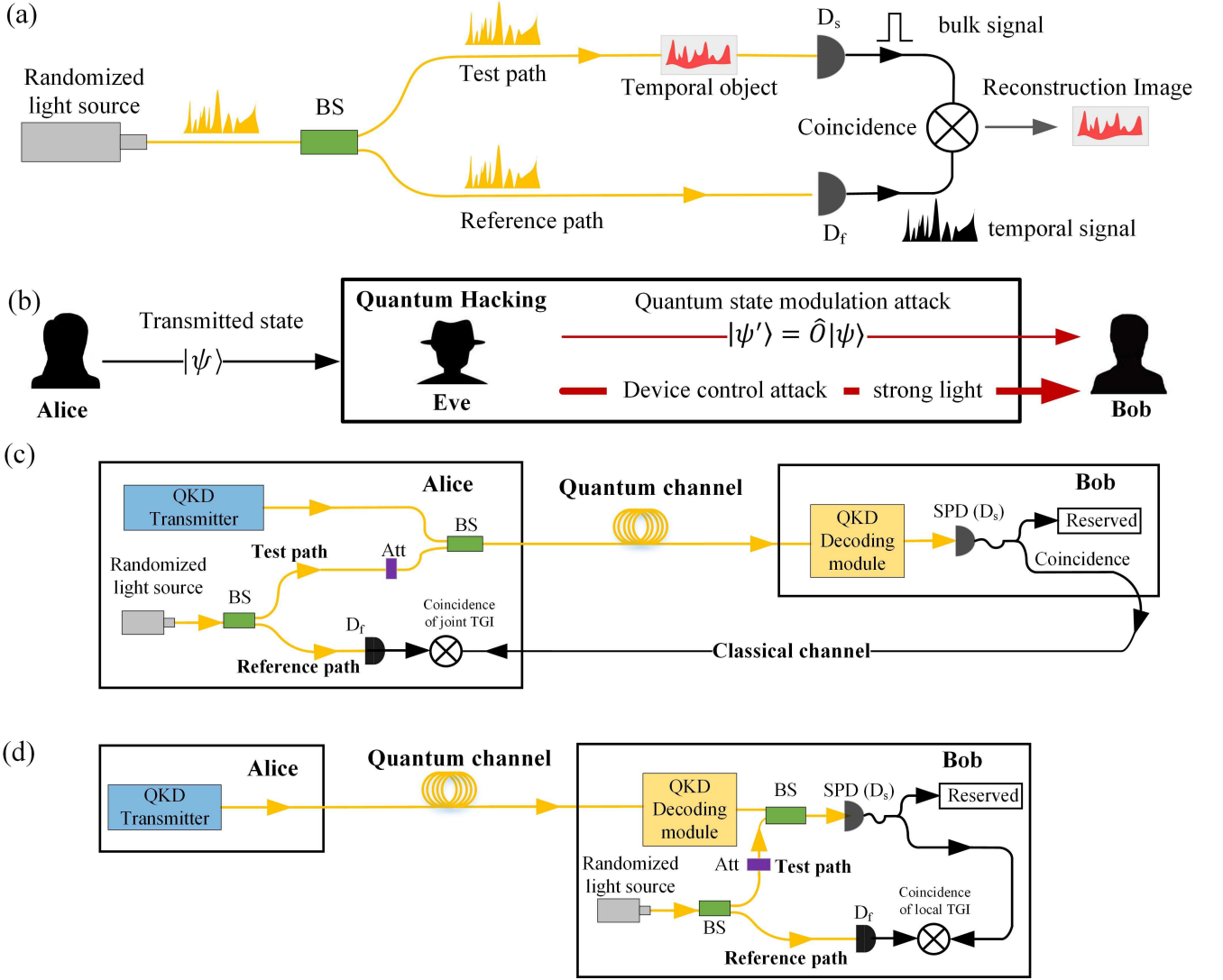}
	\caption{Concept diagrams of quantum hacking and TGI. (a) Schematic setup of temporal ghost imaging (TGI). (b) Concept diagram of quantum hacking, where Alice and Bob denote the legitimate users' transmitter and receiver, respectively. Eve denotes the eavesdropper. (c) and (d) are the concept diagrams to perceive quantum hacking with joint and local TGI methods, respectively. BS: beam splitter; Att: attenuator; SPD: single-photon detector.}
	\label{fig:concept_200331}
\end{figure*}

TGI \cite{Setala2010,Shirai2010,Ryczkowski2016} is the time-domain version of ghost imaging \cite{Pittman1995, Bennink2002,Erkmen2010,Shapiro2012}. The intrinsic mechanism of TGI is the space-time duality, which is illustrated in Fig. \ref{fig:concept_200331}(a). The randomized light source is divided into the reference path and the test path. A temporal object is located on the test path and detected by a slow detector ($D_s$), whose temporal resolution is coarse or even absent to be a bulk detector. A fast detector ($D_f$) with a high-bandwidth is placed on the reference path, and the high-resolution images of the temporal object can then be reconstructed by calculating the correlation functions of the detecting results of $D_s$ and $D_f$. The correlation function of TGI is the convolution of these two paths \cite{Shirai2010}
\begin{equation}
M(t)=\langle\Delta I_{ref}(t)\Delta I_{test}\rangle_N\propto m(t')\delta(t'-\frac{t}{s})=m(\frac{t}{s})
\label{eq:TGIfunction}
\end{equation}
where $\langle\rangle_N$ is the ensemble average over $N$ times of synchronized measurements, $\Delta I_{ref}(t)=I_{ref}(t)-\langle I_{ref}(t)\rangle_N$ and $\Delta I_{test}=I_{test}-\langle I_{test}\rangle_N$ are the intensity fluctuations of the reference and test paths, respectively. The subscript \textit{test} and \textit{ref} denote the measurement results of $D_s$ and $D_f$, respectively. $m(t')$ is the temporal object, $\delta(t'+t/s)$ is the Dirac function, and $s$ is the magnified factor \cite{Setala2010}. The magnification factor becomes $s=1$ for lensless TGI. The Dirac function $\delta(t'+t/s)$ denotes that the temporal resolution of the object only depends on the detection resolution of the reference path.

According to Eq. (\ref{eq:TGIfunction}), the temporal resolution of the reconstruction image $M(t)$ is determined by the coherence time of the randomized light source ($\tau_l$) and the temporal resolution of $D_f$, which is independent of the temporal resolution of the detector after the object. These features make TGI a potential technique for many applications, such as computational temporal imaging \cite{Devaux2016}, optical secure imaging \cite{Pan2017, Yao2018}, detection efficiency evaluation of SPDs \cite{Wu2019}, and wavelength-conversion imaging \cite{Wuh2019}.

\section{Principle of quantum hacking perceiving}
\label{Principle of quantum hacking perceiving}

Quantum hackers always try to conceal their tracks while changing the characteristics of the quantum states or quantum systems. For example, SPDs usually work in gated mode to suppress noise from ultra-weak photoelectric signals. The final detecting results of the SPDs only indicate click or not, and the precise temporal information within the tagging windows is erased. Therefore, Eve may conceal the corresponding changes due to the lack of temporal details of final detection signals. In the principle of TGI, we can treat the quantum channel and quantum system as temporal objects and can reconstruct their temporal characteristics without requiring any temporal resolution of the detector in the test path. It means that we can use TGI to reveal any behavior changing the temporal characteristics of the quantum signal, quantum channel, and quantum system without modifying the QKD system itself. 

From the perspective of exiting signals and the system response, we can classify quantum hacking into two major types (see Fig. \ref{fig:concept_200331}(b)). The first type of quantum hacking modulates transmitted quantum states as the excitation signals while avoiding detecting them directly \cite{Makarov2005a,Qi2005,Zhao2008}. The second type of quantum hacking takes an intercept-and-resend strategy by controlling of the behaviors of quantum devices in the system \cite{Lydersen2010,Wiechers2011,Gerhardt2011,Bugge2014}. 

Time-shift attack \cite{Zhao2008,Qi2005} is a representative instance of the first type, which eavesdrops the secure key by introducing additional time delays $\Delta t$ to the transmitted states (see \textbf{APPENDIX B} for more details). Taking the temporal detection efficiency of the SPD as the temporal object $m(t)$ and considering the relative time delay, the reconstruction image of Eq. (\ref{eq:TGIfunction}) becomes
\begin{equation}
M(t)\propto m(t-\Delta t)
\label{eq:time-shiftattack}
\end{equation}
where we assume a lensless TGI and have set the magnification factor $s=1$. Eq. (\ref{eq:time-shiftattack}) means the time-shift of the reconstructed temporal image and the attack behavior will be revealed.

A typical attack of the second type of quantum hacking is the blinding attack, which can entirely control the detection results of the SPDs by injecting intense laser beams into them(see \textbf{APPENDIX C} for more details). This type of quantum hacking is the most dangerous to a prepare-and-measure QKD system, such as the BB84 QKD system. If Eve blinds all detecting rounds of Bob, the attack will lead to a zero detection efficiency of the SPD at single-photon level, which means
\begin{equation}
M(t)\propto m(t)\equiv0
\label{eq:blindingattack}
\end{equation} 
Thus the reconstruction image will be nothing but noise, and the abnormal result is a clear alarm signal against the blinding attack. 

According to theoretical analysis above, we have designed two TGI schemes to perceive attacks against the QKD system. The first scheme is the joint TGI (Fig. \ref{fig:concept_200331}(c)), where the transmitter (Alice) and the receiver (Bob) separately keep different components of a TGI system and perform the joint measurement. Alice preserves the randomized light source, the reference path, and the fast detector. The test path includes the quantum channel and Bob's detection module. The light intensity of the test path is attenuated into single-photon level before transmitted to Bob. The second scheme is called local TGI (Fig. \ref{fig:concept_200331}(d)), which is solely preserved and executed by Bob. The test path of the local TGI is the detection module of Bob.

According to Eq. (\ref{eq:time-shiftattack}), the joint TGI is available to monitor the temporal distortion of the signals transmitted through the quantum channel. The local TGI is typically performed by Bob to evaluate the temporal responding characteristics of the receiver itself. According to Eq. (\ref{eq:blindingattack}), local TGI will perceive blinding attack if Eve blinds all detecting round of Bob. However, if Eve changes the attacking probability in a lossy quantum channel, the model will be more complicated. We will next prove that the local TGI is still valid for perceiving probabilistic blinding attacking. 

\section{Blinding hacking perceiving with lossy channel}
\label{Blinding hacking perceiving with lossy channel}

In a lossy quantum channel, which is the actual situation for QKD, the optimal strategy for Eve is to blind the SPDs for a small fraction rather than the total detecting rounds according to the transmission efficiency of the quantum channel to make sure that the counting rate with or without attack are identical. We will prove that TGI monitoring method is effective in this situation following.

We denote the response of the SPD as $I_{test}\in\{0,1\}$, where 0 and 1 indicate that there is no and one click of the SPD in a detecting window, respectively. If their is no attack to the QKD system, the SPD clicks are triggered by Bob's TGI signals, Alice's quantum signals, and the dark count of the SPD, which are labeled as $I_{tb}$, $I_{ta}$, and $I_{td}$, respectively. Then $I_{test}=1-(1-I_{tb})(1-I_{ta})(1-I_{td})$. If we substitute $I_{test}$ into Eq. (\ref{eq:TGIfunction}), we obtain the amended TGI correlation function without attack as 
\begin{equation}
\begin{aligned}
M_1(t)=(1-\langle I_{td}\rangle)(1-\langle I_{ta}\rangle)\langle\Delta I_{ref}(t)\Delta I_{tb}\rangle
\end{aligned}
\label{eq:MtAlice}
\end{equation}
where, $\langle\Delta I_{ref}(t)\Delta I_{tb}\rangle$ is the reconstructed image of the TGI when there is no quantum signal from Alice or dark count of the SPD. Equation (\ref{eq:MtAlice}) shows that the quantum signals or dark counts decrease the amplitude of the reconstructed image linearly.

In a blinding attack scenario with an intercept and resend strategy, Eve can resend fake states with a certain probability, along with the control signals to manipulate the responses of the SPDs. We denote the attack behavior using two parameters $I_{te0}$ and $I_{te1}$. Under the blinding attack, if Bob chooses a different measurement basis from Eve, the attacking light splits into two paths and the detection current of Bob's SPD will be smaller than the threshold. Thus, there will be no click at all, which will be denoted as $I_{te0}=1, I_{te1}=0$. Oppositely, if Bob chooses the same basis with that of Eve, the attacking light will incident into a single path and lead to a detection current above the threshold. In this case, there will certainly be a click event from Bob's SPD, which will be denoted as $I_{te0}=0, I_{te1}=1$. Obviously, $I_{te0}=I_{te1}=0$ when Eve doesn't attack Bob's SPD. Then the response of the SPD becomes $I_{test}^{blind}=[1-(1-I_{tb})(1-I_{te1})(1-I_{td})](1-I_{te0})$. By substituting $I_{test}$ with $I_{test}^{blind}$ in Eq. (\ref{eq:TGIfunction}) and keeping $I_{te0}*I_{te1}=0$ in mind, we obtain the correlation function under attack as 
\begin{equation}
M_2(t)=(1-\langle I_{td}\rangle)(1-\langle I_{te1}\rangle-\langle I_{te0}\rangle)\langle\Delta I_{ref}(t)\Delta I_{tb}\rangle
\label{eq:MtEve}
\end{equation}
See \textbf{APPENDIX D} for proofs of Eqs. (\ref{eq:MtAlice})-(\ref{eq:MtEve}). The difference of these scenarios with and without attack can be evaluated using the differential image 
\begin{equation}
\begin{aligned}
\Delta M(t)=&M_1(t)-M_2(t)\\
=&(1-\langle I_{td}\rangle)(\langle I_{te1}\rangle+\langle I_{te0}\rangle-\langle I_{ta}\rangle)\langle\Delta I_{ref}(t)\Delta I_{tb}\rangle
\end{aligned}
\label{eq:deltaMt}
\end{equation}
In order to cheat Bob, Eve prefers to make $\langle I_{te1}\rangle=\langle I_{ta}\rangle$ so that the count rate seems to be normal. Since the average dark count rates of the SPDs usually satisfy $\langle I_{td}\rangle\ll 1$, we can simplify Eq. (\ref{eq:deltaMt}) to
\begin{equation}
\Delta M(t)\approx \langle I_{te0}\rangle\langle\Delta I_{ref}(t)\Delta I_{tb}\rangle
\label{eq:deltaMtapprox}
\end{equation}
Equation (\ref{eq:deltaMtapprox}) shows that the differential image is proportional to the original TGI $\langle\Delta I_{ref}(t)\Delta I_{tb}\rangle$ while the amplitude is proportional to the average attack probability. It should be mentioned that Bob cannot distinguish legitimate user from the hacker directly and hence could not use Eq. (\ref{eq:deltaMtapprox}) directly to perceive attacks. However, Bob could obtain the measured image by his setup according to Eq. (\ref{eq:TGIfunction}). This image will be equivalent to Eq. \ref{eq:MtAlice} ( or Eq. (\ref{eq:MtEve})) when there is (or no) attack. And he could also obtain $\langle\Delta I_{ref}(t)\Delta I_{tb}\rangle$, $\langle I_{ta}\rangle$ (or $\langle I_{te1}\rangle$, anyway, $\langle I_{te1}\rangle=\langle I_{ta}\rangle$) and $\langle I_{td}\rangle$ directly by his local TGI and QKD module. That means Bob could also reconstruct a predicted legitimate image according to Eq. (\ref{eq:MtAlice}). The differential image between the measured and predicted legitimate images will be noise only when there is no attack ($I_{te0}=I_{te1}=0$). Once Eve launches Blinding attack, the probability of case "$I_{te0}=1, I_{te1}=0$" will never be zero, as Bob's bases chosen is independent to Eve. That means $\langle I_{te0}\rangle\neq 0$ and the reconstruction differential image will be proportional to Eq. (\ref{eq:TGIfunction}). Hence, Eve's attack will be clearly revealed. 

Therefore, the combination of the joint and the local TGI can deal with the two types of quantum hacking attacks mentioned above. To meet the security criteria of QKD, Alice and Bob can randomly choose to perform QKD sessions or TGI sessions independently with a prearranged duty cycle and disclose the sessions they performed during the post-processing of QKD. It is worth noting that they do not need to announce their detection results except for Bob's detecting results during the joint TGI. The monitoring system works as follows: 

1. Alice and Bob randomly execute the QKD or TGI procedures; 

2. After sufficient rounds, Alice and Bob announce the rounds that execute the joint and local TGIs, respectively, and abandon the rounds that both TGIs are executed; 

3. Bob sends the click results of the SPDs of the preserved joint TGI rounds to Alice; 

4. Alice and Bob reconstruct the temporal detection efficiencies of the SPDs according to the joint and local monitoring results, respectively, and reveal the attack behaviors.

\section{Experimental demonstration}
\label{Experimental demonstration}

\begin{figure*}
	\centering
	\includegraphics[width=\textwidth]{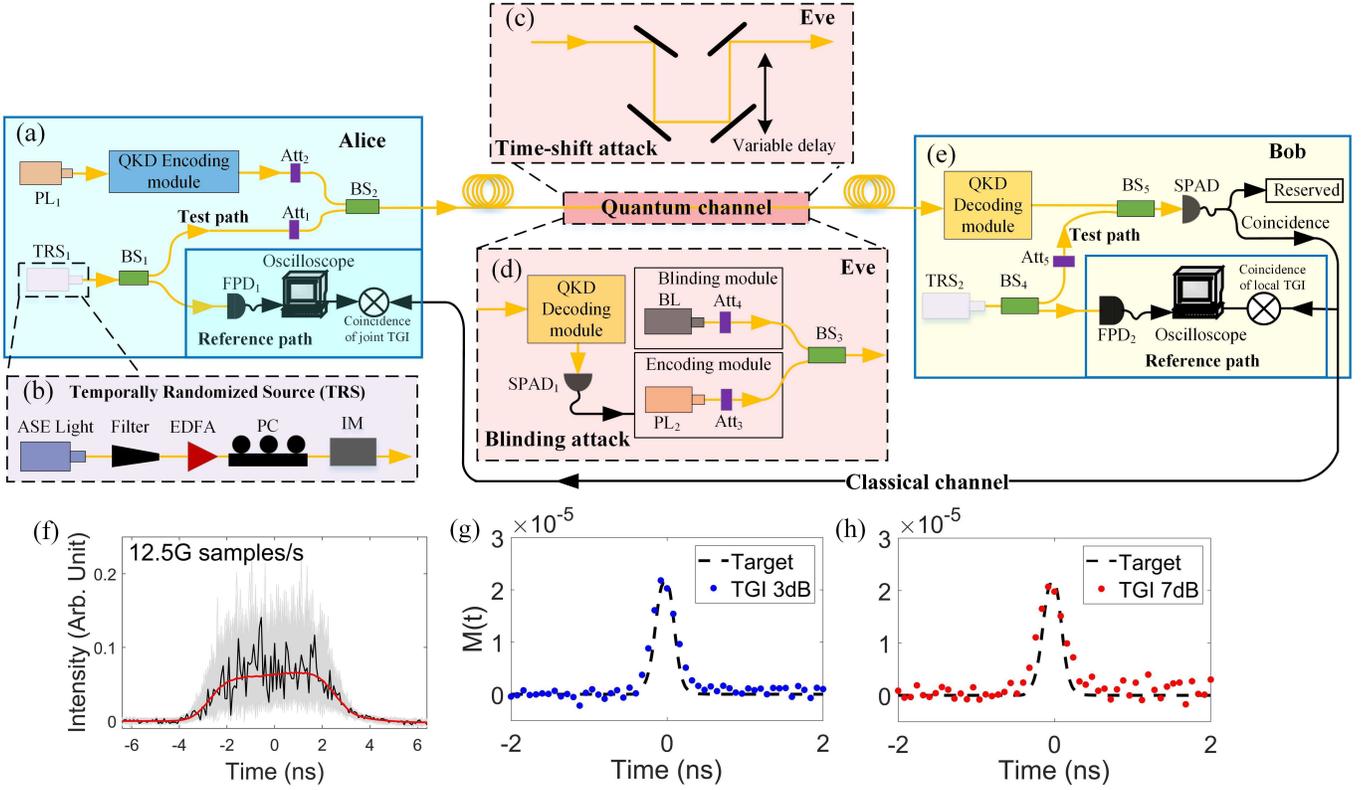}
	\caption{Proof-of-principle experimental setup of the TGI monitoring method for quantum hacking perceiving. (a) The experimental setup of the transmitter Alice. The QKD transmitter consist of a pulsed laser (PL${}_1$) and the encoding module. Alice reserve the temporally randomized source (TRS${}_1$) and the reference path of the joint TGI. The test path consists of a fast photodiode (FPD${}_1$) and a real-time oscilloscope with a bandwidth of 25 GHz and 12.5GHz, respectively. (b) The setup of the TRS, which consists of an amplified spontaneous emission (ASE), a 50-GHz-passband optical filter, an erbium-doped fiber amplifier (EDFA), a polarization controller (PC) and an intensity modulator (IM). (c)-(d), The proof-of-principle setup of the time-shift attack and the blinding attack. (e) Bob's setup, which includes the QKD receiver and the local TGI monitoring module. The QKD receiver, including the decoding and the detection modules, is involved in the test path of the joint TGI. (f) The real-time intensity fluctuation of the TRS. The black line is the intensity fluctuation within a single measurement window. The gray background and the red line are the overlapping results and the average value of 5000 measurement events, respectively. (g) and (h) Reconstruction images of the joint TGI without quantum hacking under 3 dB and 7 dB equivalent channel loss, respectively. The black dashed line is the target image (the temporal object) of the SPAD.PL, pulsed laser; BS, beam splitter; Att, attenuator; BL: blinding laser; SPAD, single-photon avalanche detector.}
	\label{fig:setup_quantum hacking perceiving_200331}
\end{figure*}

\subsection{The proof-of-principle demo system}
\label{The proof-of-principle demo system}

The proof-of-principle experimental setup to perceive quantum hacking using the TGI monitoring method is shown in Fig. \ref{fig:setup_quantum hacking perceiving_200331}. The QKD transmitter in Alice (Fig. \ref{fig:setup_quantum hacking perceiving_200331}(a)) consists of a pulsed laser (PL${}_1$) and a QKD encoding module. Alice also keeps a temporally randomized source (TRS) and the reference path of the joint TGI (Fig. \ref{fig:setup_quantum hacking perceiving_200331}(b)). The TRS of the joint TGI (TRS${}_1$) is divided into the reference path and the test path by a beam splitter (BS${}_1$). The reference light is detected by a fast photodiode (FPD${}_1$) and an oscilloscope with the bandwidth of 25GHZ and 12.5GHz, respectively. 

Alice can randomly execute QKD or the joint TGI procedures for security evaluation. Eve may attack the system by launching the time-shift attack (Fig. \ref{fig:setup_quantum hacking perceiving_200331}(c)) or blinding attack (Fig. \ref{fig:setup_quantum hacking perceiving_200331}(d)) using a variable controlled optical delayer or a intercept-and-resend unit. Therefore, Bob may receive the legitimate quantum photons, the TGI monitoring signals, or the attack light from the quantum channel. Bob has a QKD decoder, including the photon detecting modules, and the local TGI monitoring unit (Fig. \ref{fig:setup_quantum hacking perceiving_200331}(e)). The test path of the local TGI is merged into the SPD with the decoding signals from the quantum channel. In our experiment, we use single-photon avalanche detectors (SPADs) to detect single photons. The SPADs have no temporal resolution within a detection round. The detection results of SPADs are reserved as the raw key or TGI signals after consulting with Alice. Since the detection results of TGI are only used for security evaluating and then abandoned, the proportion of TGI sessions can be kept small to increase the key generation efficiency. 

We implement the TRS (Fig. \ref{fig:setup_quantum hacking perceiving_200331}(b)) from an amplified spontaneous emission (ASE) light, which is amplified by an erbium-doped fiber amplifier (EDFA) and then chopped into pulses using an intensity modulator (IM). It should be mentioned that an optical filter with a bandwidth of 50GHz is cascaded to the ASE source to constrain its output bandwidth in order to match the bandwidth of the fast photodiode (FPD${}_1$) and oscilloscope \cite{Ryczkowski2016}. 

Fig.\ref{fig:setup_quantum hacking perceiving_200331}(f) indicates the real-time intensity fluctuation of the TRS. The black line is the intensity fluctuation of a single measurement window. The gray background and the red line are the overlapping results and average value of 5000 measurement windows, respectively. The intensity fluctuation indicates a short effective characteristic time about 80 ps, which is the upper temporal resolution of the TGI.

\subsection{Defending time-shift attack}
\label{Defending time-shift attack}

We execute the joint TGI in our experiment to defend the time-shift attack, and the influence of QKD and local TGI can be eliminated during the announcement step. The intensity of TGI photons sent to Bob is about $\mu_t\simeq 0.6$ within every detecting window. We firstly reconstruct the temporal images of Bob's SPAD in the test path of TGI without any attack. Figures \ref{fig:setup_quantum hacking perceiving_200331}(g)-\ref{fig:setup_quantum hacking perceiving_200331}(h) gives the reconstructed images through the quantum channel with 3 dB and 7 dB transmission loss, respectively, with a sample block size of $N=5\times10^6$. We can see that the statistical fluctuation of the image under 7 dB loss is slightly higher because the effective counting rate of the SPAD is relatively low, which can be decreased by increasing the sample size.

\begin{figure*}
	\centering
	\includegraphics[width=\textwidth]{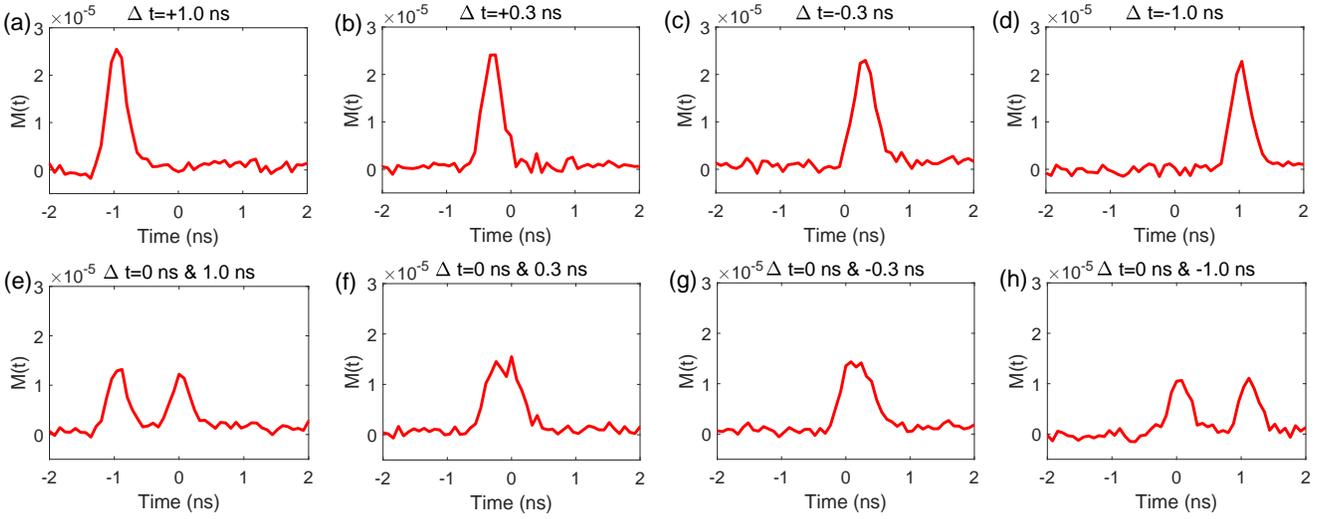}
	\caption{Reconstruction images of the joint TGI against time-shift attack. (a)-(d) Images with different time-delay values. (e)-(h), Images under the time-shift attack, where the time delay is randomly chosen between two values.}
	\label{fig:timeshiftattack}
\end{figure*}

A variable optical delayer is added via the 3-dB-loss quantum channel to simulate the time-shift attack by Eve. Figures \ref{fig:timeshiftattack}(a)-\ref{fig:timeshiftattack}(d) show the reconstructed temporal images under different time delays ($\Delta t=1.0$ ns, 0.3 ns, -0.3 ns, -1.0 ns), which exhibit that the time-shift attack can be regarded as a translation operator to a temporal object (such as the SPAD) and can be monitored using TGI. 

The reconstructed images under time-shift attack randomly switching between two delay values are shown in Figs. \ref{fig:timeshiftattack}(e)-\ref{fig:timeshiftattack}(h). In these situations, the reconstructed images (the red lines) become the superposition of that under a single time delay. Comparing with that under no attack (Figs. \ref{fig:setup_quantum hacking perceiving_200331}(g)-\ref{fig:setup_quantum hacking perceiving_200331}(h)), the time-shift attack can be obviously revealed even when the shifting time is as short as 0.30 ns.

\subsection{Defending blinding attack}
\label{Defending blinding attack}

\begin{figure*}
	\centering
	\includegraphics[width=0.9\textwidth]{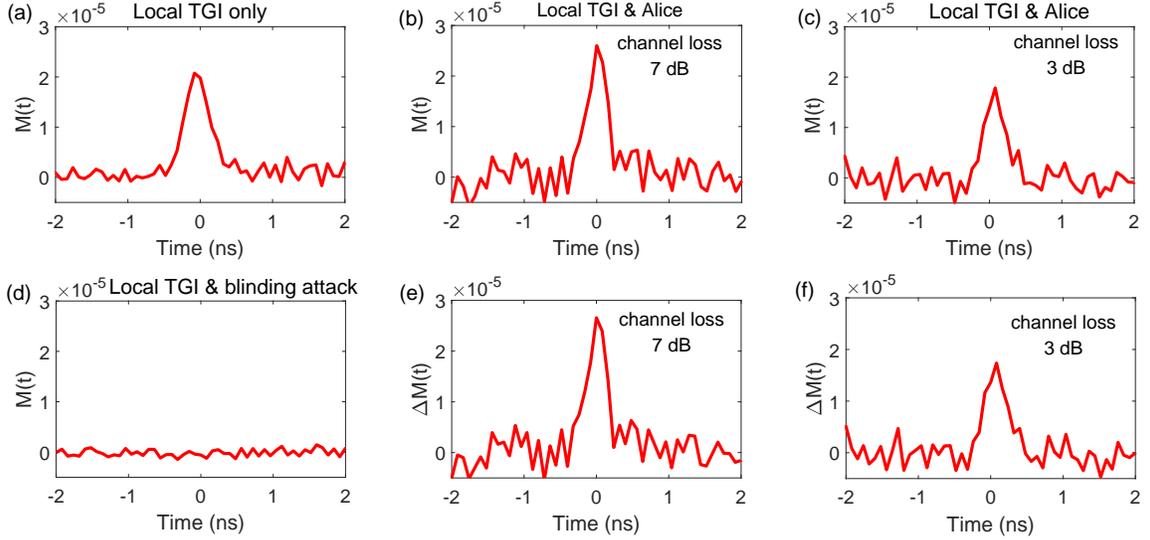}
	\caption{The monitoring results of local TGI against blinding attack. (a) The reconstructed image of the local TGI without signals from the quantum channel. (b)-(c) Reconstructed image of the local TGI with Alice sending signals over 3 dB and 7 dB loss quantum channels, respectively. (d) Reconstructed image under blinding attack. (e)-(f) Differential images over quantum channels with 3 dB and 7 dB transmission loss, respectively, where Eve attacks all rounds.}
	\label{fig:blindingattack}
\end{figure*}

We use the local TGI in Bob to defend the blinding attack. According to Eqs. (\ref{eq:MtAlice})-(\ref{eq:MtEve}), the signals from Alice and Eve are noise to the local TGI in Bob. We first give the reconstructed images with no attack. According to Eq. (\ref{eq:MtAlice}), the reconstructed image is not sensitive to counts triggered by Alice as long as $\langle I_{ta}\rangle$ is much smaller than 1. To show this merit, we set the average photon number of the transmitted quantum state sent by Alice being $\mu_{a}=0.5$, and the effective photon number of the local TGI being $\mu_t\simeq0.12$ within each detection window so that $\langle I_{ta}\rangle$ is much larger than $\langle I_{tb}\rangle$. 

Figures \ref{fig:blindingattack}(a) and \ref{fig:blindingattack}(b)-\ref{fig:blindingattack}(c) show the local TGI images without and with QKD, respectively. The average count rates triggered by the local TGI is $\langle I_{tb}\rangle=0.01$, and that triggered by the quantum photons from Alice ($\langle I_{ta}\rangle$) are about 0.025 and 0.050 per round, respectively, over 7dB and 3dB loss channels. As $\langle I_{ta}\rangle$ and the dark count rate of the SPAD is much smaller than 1 ($\langle I_{td}\rangle\approx5\times10^{-5}$), Eq. (\ref{eq:MtAlice}) indicates that the difference between these reconstructed images will approximate to each other. The experiment results in Figs. \ref{fig:blindingattack}(a)-\ref{fig:blindingattack}(c) show that there are no significant difference between the reconstructed images in the situations above when taking the statistical fluctuation into account. Figure \ref{fig:blindingattack}(d) is the result when Eve blinds Bob's SPAD, which only has statistical fluctuation and fits Eq. (\ref{eq:blindingattack}). Figures \ref{fig:blindingattack}(e)-\ref{fig:blindingattack}(f) are the differential images over 7dB and 3dB loss channels, respectively. The differential images clearly reveal the existence of blinding attack, because there should be nothing but statistical fluctuations when the SPAD runs normally according to Eq. (\ref{eq:deltaMtapprox}).

\subsection{Defending blinding attack over lossy channel}
\label{Defending blinding attack over lossy channel}

\begin{figure*}
	\centering
	\includegraphics[width=\textwidth]{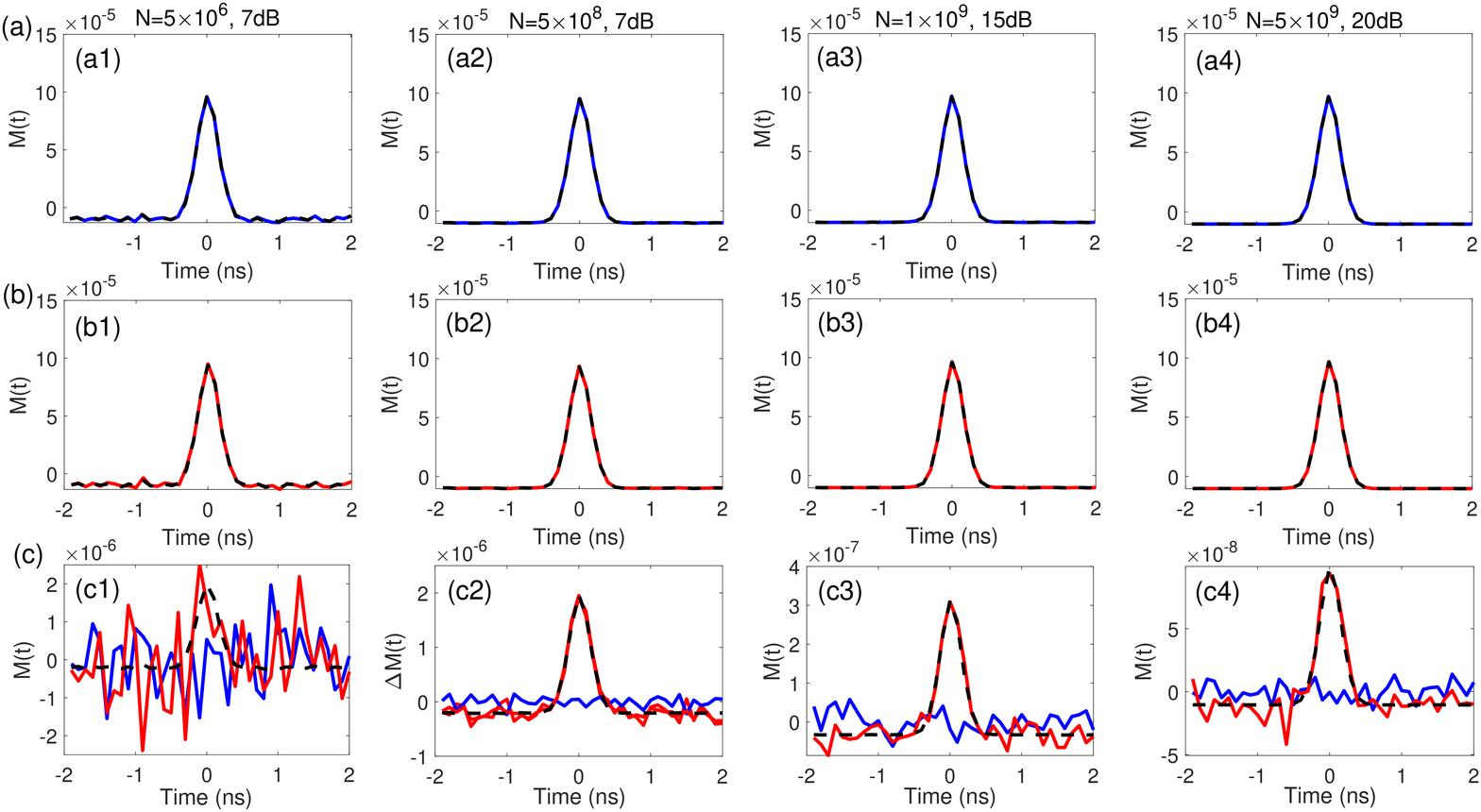}
	\caption{The simulation results of local TGI against blinding attack over lossy channels. (a) Reconstructed images of the local TGI when Alice sends signals over lossy channels. (b) Reconstruction images of the local TGI when Eve blinds a small fraction of detection windows.(c) The differential images between the predicted (Eq. (\ref{eq:MtAlice})) and measured (Eq. (\ref{eq:TGIfunction})) images with (red lines) and without (blue lines) attack, respectively. The black lines are the asymptotic differential images, according to Eq. (\ref{eq:deltaMtapprox}). $N$ is the size of the statistical sample, and 7 dB means the channel loss is 7 dB.}
	\label{fig:blindingattacksimulation}
\end{figure*}

In the experiments above, Eve blinds every detection window of Bob's SPAD. However, in a practical high-loss quantum channel, a better strategy for  Eve is to attack a portion of the detecting windows of the system. According to Eq. (\ref{eq:deltaMtapprox}), the differential image of local TGI with and without blinding attack is proportional to the average attacking probability. Though $\langle I_{te0}\rangle$ decreases as the channel loss increases, it can never be zero and the attack will be revealed clearly. Larger sample size is required to reveal the attack for a higher lossy channel for practical QKD sessions.

Figure \ref{fig:blindingattacksimulation} gives the simulation results of the local TGI against blinding attack over lossy channels. The simulation parameters are the same as those of the experimental system except for the TRS. The intensity of the TRS here is uniformly distributed between 0 and 1, the time resolution is about 80 ps and $\langle I_{tb}\rangle\simeq 0.050$. The first row (Figs. \ref{fig:blindingattacksimulation}(a1)-\ref{fig:blindingattacksimulation}(a4)) are the predicted (Eq. (\ref{eq:MtAlice}), black dashed lines) and "measured" (Eq. (\ref{eq:TGIfunction}), blue lines) images of the local TGI without attack. The second row (Figs. \ref{fig:blindingattacksimulation}(b1)-\ref{fig:blindingattacksimulation}(b4)) are the corresponding predicted (black lines) and "measured" (red lines) images with blinding attack, where $\langle I_{te1}\rangle=\langle I_{te0}\rangle=1-exp(-\alpha\mu_a\eta)$, $\alpha$ is the channel loss and $\eta$ is the detection efficiency of the SPADs. The blue and red lines of the third row (Figs. \ref{fig:blindingattacksimulation}(c1)-(c4)) show the differential images between the predicted and "measured" reconstructed images without (Figs. \ref{fig:blindingattacksimulation}(a1)-(a4)) and with (Figs. \ref{fig:blindingattacksimulation}(b1)-(b4)) attack, respectively. And the black lines of the third row are the asymptotic differential images constructed by Eq. (\ref{eq:deltaMtapprox}) directly without considering statistical fluctuation. The consistency of the predicted and "measured" images is consistent to the theoretical expectations. For a lossy channel, a sample size of $N=5\times10^6$ is not sufficient to reconstruct a high signal-to-noise differential image (see the red line in Fig. \ref{fig:blindingattacksimulation}(c1)). Though the memory size of our experiment limits the sample size acquired, a high-quality differential image can be achieved when the sample size expands to $5\times10^8$ (Fig. \ref{fig:blindingattacksimulation}c2), and a sample size of  $5\times10^9$ is sufficient to support a high-quality differential image through a channel with 20dB loss (Fig. \ref{fig:blindingattacksimulation}c4).

\section{Discussion and conclusion}
\label{Discussion and conclusion}

According to Fig. \ref{fig:blindingattacksimulation}, the sample size required to perceiving blinding attack over high-loss channel is relatively large in the demo system. However, by combining state-of-the-art techniques, such as quantum, compressive and differential ghost imaging methods \cite{Erkmen2010,Shapiro2012,Katz2009,O-Oka2017,Ferri2010}, the sample size required will be reduced by several orders, which means that the TGI system will be easier to implement and has the potential to apply in real-time monitoring of practical QKD system. 

For the joint TGI, the identity of the TRS and the QKD light source is essential for defending against eavesdropping. The bandwidth of the filtered TRS is 50 GHz in our experiment, which is at the same level as the light sources commonly used in practical QKD systems \cite{Boaron2018,Dynes2018}. The ASE or LED light is not only a perfect temporally randomized source (TRS), but also a potential candidate for economical and portable QKD devices \cite{Chun2018}. Therefore, the method will be effective by using the same light source, encoding and decoding modules in both the QKD and TGI procedures. Although we only use one SPD in this proof-of-principle experiment, the method is untroubled to be applied to a conventional QKD system with multiple SPDs. 

Usually, quantum hacking against the QKD system may be probabilistic and dynamic, while TGI requires the temporal object to be "stationary" during a reconstruction period. However, according to Eq. (\ref{eq:deltaMtapprox}), the TGI monitoring does not require a stationary attack from Eve. The monitoring is effective if only $\langle I_{te0}\rangle\neq 0$, which always stands once Eve blinds Bob's SPD and no matter what the attack probability is.

In conclusion, we have proposed an effective quantum hacking perceiving method using TGI. Due to the missing temporal resolution, some of the quantum devices like SPAD act as temporal black boxes, which provides eavesdroppers chances to hide their hacking evidence. However, the TGI method makes it possible to directly image the SPADs without changing their behaviors at the single-photon level. The method we proposed makes the black box transparent and can reveal the temporal quantum hacking evidence, which provides a novel measure in quantum hacking perceiving. As has been tried in previous works \cite{Maroy2017,Pinheiro2018}, one meaningful study in the future is to integrate the imaging perceiving method into the security proof of the system. Moreover, to defend against the attacks in other degrees of freedom (DOFs) can be developed using the single-photon imaging technology analogously to ghost imaging.

\section*{Acknowledgments}
This work has been supported by National Key Research and Development Program of China (Grant No. 2018YFA0306400), National Natural Science Foundation of China (Grant Nos. 61627820, 61675189, 61905235, 61622506, 61822115), Anhui Initiative in Quantum Information Technologies(Grant No. AHY030000). F.-X.Wang has also been supported by China Postdoctoral Science Foundation (2019M652179).

\section*{Disclosures}
The authors declare no conflicts of interest.

\section*{Appendix A: Single-photon detector}
\label{appendix-a}

The SPADs used in the experiment are commercial products from Anhui Qasky Quantum Technology Co. Ltd. The working gating frequency of the SPAD is 10MHz. The full width at half maximum (FWHM) of the detection window is about 0.27 ns, and the peak detection efficiency is 21.4\%. The dark count rate is about $5\times 10^{-5}$ per gate. 

\section*{Appendix B: Time-shift attack}
\label{appendix-b}

\renewcommand\thefigure{S\arabic{figure}}  
\setcounter{figure}{0}  

\begin{figure*}
	\centering
	\includegraphics[width=0.85\textwidth]{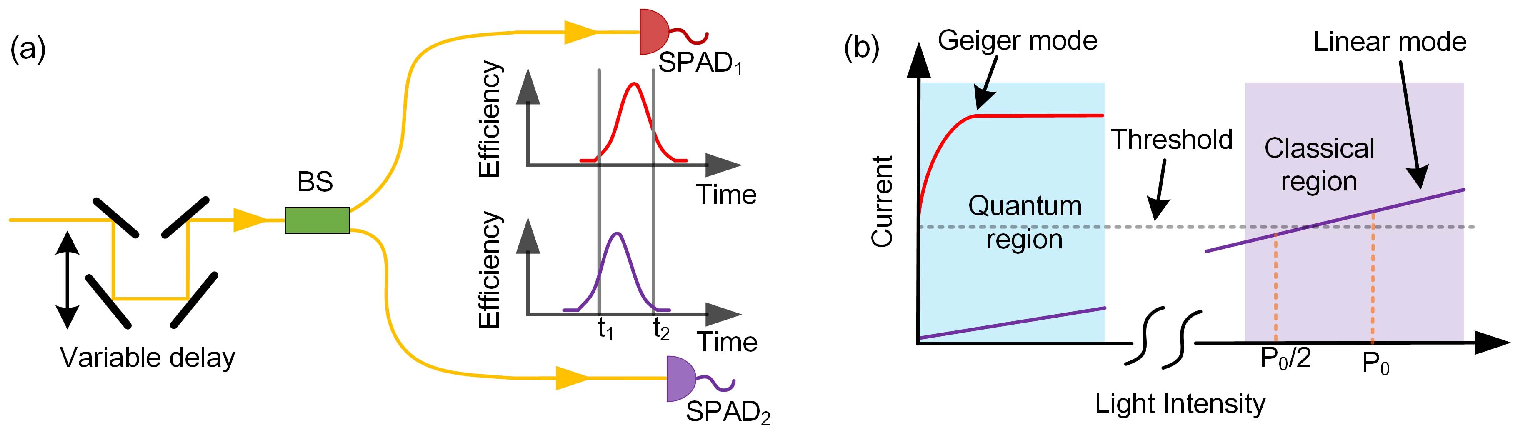}
	\caption{Principle of time-shift attack and blinding attack. \textbf{a,} The principle of time-shift attack, where temporal detection efficiency mismatch (see $t_1$ and $t_2$) between SPADs (the red and purple efficiency curves) is the key of attack. \textbf{b,} Current of the SPAD with input light intensity under Geiger mode (red curve) and linear mode (purple curve), respectively. The dashed line is the threshold, above which the SPAD will be triggered. The quantum (blue background) and classical (purple) regions represent a single-photon level (e.g., less than $10^2$ photons) and a strong intensity level (e.g., larger than $10^5$ photons) of input light intensity, respectively.}
	\label{fig:attackprinciple}
\end{figure*}

The time-shift attack exploits the efficiency mismatch of SPDs working in Geiger mode \cite{Qi2005,Zhao2008}. The attack principle of the time-shift attack is shown in Fig. \ref{fig:attackprinciple}\textbf{a}. The temporal detection efficiencies of Bob's SPADs are slightly mismatched. Eve can introduce two different time delays (corresponding to $t_1$ and $t_2$, respectively) to transmitted quantum states via the quantum channel. Then, the detection results and the time delay can be one-to-one mapped exploring the detection efficiency mismatch between SPAD${}_1$ (the red curve) and SPAD${}_2$ (the purple curve). For example, when the arrival time is $t_1$, only SPAD${}_2$ responds to the received photon. Eve will get the full information of the sifted key between Alice and Bob after Bob announced his measuring basis for every round.

\section*{Appendix C: Blinding attack}
\label{appendix-c}

Blinding attack alters the behavior of the SPD from Geiger mode to linear mode by injecting illuminating light with proper power \cite{Lydersen2010}. The current of the SPAD increases rapidly via input light intensity in Geiger mode (red curve in Fig. \ref{fig:attackprinciple}\textbf{b}). When a strong continuous-wave (CW) or a pulsed laser beam is incident into the SPAD, the SPAD will be switched to the linear mode from Geiger mode. The corresponding current in linear mode increases much slower via input light intensity (purple line in Fig. \ref{fig:attackprinciple}\textbf{b}). The corresponding current caused by single-photon-level light (the "quantum region" with a blue background) is too small to reach the discrimination threshold and could not trigger a "click" signal. That is, the SPAD is blind to the single-photon signals. However, if the input light intensity becomes strong enough (the "classical region" with a purple background), the corresponding current will be large enough to trigger a "click" signal. By exploiting the above characteristic of linear mode, Eve can control SPADs of the QKD system and acquire full information of the final key without introducing significant QBER. The strategy that Eve adopts is as follows. Eve intercepts and decodes the transmitted state as same as Bob does. Then Eve encodes and resends a strong pulsed light with power $P_0$ to Bob according to his detection results. If Bob chooses the same basis as Eve does, the pulse will inject into a single SPAD and will cause a high-level current to trigger a "click", which can be denoted by $I_{te1}=1$. If Bob chooses a measuring basis different from that of Eve, the pulse will be divided into two SPADs equally. The corresponding current will be lower than the threshold to trigger a "click", which corresponds to $I_{te1}=0$. It should be noted that when the blinding attack is taken and $I_{te1}=0$, the SPD will not respond to the single-photon-level signals from the test path of the TGI and dark count. Thus, we introduce the parameter $I_{te0}$ to differentiate it from the no-attack situation. When there is no response under blinding attack, $I_{te0}=1$. Otherwise, $I_{te0}=0$. Due to there is only one possible output from the SPD, $I_{te0}$ and $I_{te1}$ should satisfy $I_{te1}I_{te0}=0$. It means that $I_{te1}=1$ and $I_{te0}=0$ when Bob chooses the same basis as Eve does, and $I_{te1}=0$ and $I_{te0}=1$ when Bob chooses the basis different from that of Eve. The response function of the SPD becomes $I_{te1}(1-I_{te0})$. By considering the situation without blinding attack, the response function of the SPD becomes $I_{test}^{blind}=[1-(1-I_{tb})(1-I_{te1})(1-I_{td})](1-I_{te0})$.

The intensity of the CW blinding laser (BL) and the pulsed laser (PL${}_2$ in Fig. \textcolor{blue}{2}) used in the demo experiment is 15.4 $\mu$W and 7.46 $\mu$W, respectively.

\section*{Appendix D: Proof of Eqs. (4) and (5)} 
\label{appendix-d}

When there is no attack and Alice sends signals to Bob, the response function of Bob's SPD is $I_{test}=1-(1-I_{tb})(1-I_{ta})(1-I_{td})$, where $I_{tb},I_{ta}, I_{td}\in\{0,1\}$. By substituting $I_{test}$ above into Eq. (\textcolor{blue}{1}), the reconstruction image of the local TGI becomes
\begin{equation}
\begin{aligned}
&\quad M_1(t)=\langle\Delta I_{ref}\Delta I_{test}\rangle\\
&=\langle I_{ref}I_{test}\rangle-\langle I_{ref}\rangle \langle I_{test}\rangle\\
&=Cov(I_{ref},I_{tb})+Cov(I_{ref},I_{ta})+Cov(I_{ref},I_{td})\\
&\quad-\langle I_{ta}\rangle Cov(I_{ref},I_{tb})-\langle I_{td}\rangle Cov(I_{ref},I_{tb})\\
&\quad+\langle I_{ta}I_{td}\rangle Cov(I_{ref},I_{tb})-Cov(I_{ref},I_{ta}I_{td})\\
&=(1-\langle I_{td}\rangle)(1-\langle I_{ta}\rangle)\langle\Delta I_{ref}(t)\Delta I_{tb}\rangle
\end{aligned}
\label{eq:MtAlicederivation}
\end{equation}
where $Cov(I_{ref},I_{tb})=\langle I_{ref}I_{tb}\rangle$ and we have adopted $\langle I_{ref}I_{ta}I_{td}\rangle=\langle I_{ref}\rangle\langle I_{ta}\rangle\langle I_{td}\rangle$ and $Cov(I_{ref},I_{ta})=Cov(I_{ref},I_{td})=0$, as $I_{ref}$, $I_{ta}$ and $I_{td}$ are independent of each other. Eq. (\textcolor{blue}{5}) can be derived analogously. 

The derivation has not consider statistical fluctuation of all terms. As the differential image is proportional to the average attack probability $I_{te0}$, these eliminated terms of the formula, such as $Cov(I_{ref},I_{ta})$, actually fluctuate around zero and may decrease the differential imaging quality significantly when $I_{te0}$ is relative small. But the fluctuation will decrease to the acceptable level when $N$ becomes larger, which is strongly supported by Figs.\textcolor{blue}{5}\textbf{c1}-\textcolor{blue}{5}\textbf{c2} of the maintext.

\end{document}